\documentclass[twocolumn]{aastex62}

\usepackage{amsmath,amssymb}
\usepackage{epstopdf}
\usepackage{ulem}
\usepackage{multirow}
\usepackage{soul}

\shorttitle{Electron heating at SNR shocks}
\shortauthors{Bohdan et al.}

\newcommand{\mpo}{\textcolor{black}}
\newcommand{\ab}{\textcolor{black}}
\newcommand{\jn}{\textcolor{black}}

\newcommand{\pjm}{\textcolor{black}}

\def\ee{\end{equation}}
\def\be{\begin{equation}}

\newcommand{\omci}{\Omega_\mathrm{i}}

\newcommand{\ms}{M_\mathrm{s}}

\newcommand{\ma}{M_\mathrm{A}}
\newcommand{\mi}{m_\mathrm{i}}
\newcommand{\me}{m_\mathrm{e}}
\newcommand{\lse}{\lambda_\mathrm{se}}
\newcommand{\lsi}{\lambda_\mathrm{si}}
\newcommand{\vsh}{v_\mathrm{sh}}

\newcommand{\wtot}{W_\mathrm{tot}}

\newcommand{\ti}{T_\mathrm{i}}
\newcommand{\te}{T_\mathrm{e}}

\begin{document}

\title{Kinetic simulation of nonrelativistic perpendicular shocks of young supernova remnants. IV. Electron heating.}

\correspondingauthor{Artem Bohdan}
\email{artem.bohdan@desy.de}

\author[0000-0002-5680-0766]{Artem Bohdan}
\affil{DESY, 15738 Zeuthen, Germany}

\author{Martin Pohl}
\affil{DESY, 15738 Zeuthen, Germany}
\affil{Institute of Physics and Astronomy, University of Potsdam, 14476 Potsdam, Germany}

\author{Jacek Niemiec}
\affil{Institute of Nuclear Physics Polish Academy of Sciences, PL-31342 Krakow, Poland}

\author{Paul Morris}
\affil{DESY, 15738 Zeuthen, Germany}

\author{Yosuke Matsumoto}
\affil{Department of Physics, Chiba University, 1-33 Yayoi-cho, Inage-ku, Chiba 263-8522, Japan}

\author{Takanobu Amano}
\affil{Department of Earth and Planetary Science, the University of Tokyo, 7-3-1 Hongo, Bunkyo-ku, Tokyo 113-0033, Japan}

\author{Masahiro Hoshino}
\affil{Department of Earth and Planetary Science, the University of Tokyo, 7-3-1 Hongo, Bunkyo-ku, Tokyo 113-0033, Japan}

\begin{abstract}

High Mach number collisionless shocks are found in planetary systems and supernova remnants (SNRs). Electrons are heated at these shocks to the temperature well above the Rankine-Hugoniot prediction. However processes responsible for electron heating are still not well understood. We use a set of large-scale Particle-In-Cell simulations of non-relativistic shocks in high Mach number regime to clarify the electron heating processes.
The physics of these shocks is defined by ion reflection at the shock ramp. Further interaction of the reflected ions and the upstream plasma excites electrostatic Buneman and two-stream ion-ion Weibel instabilities. 
Electrons are heated via shock surfing acceleration, the shock potential, magnetic reconnection, stochastic Fermi scattering and the shock compression. The main contributor is the shock potential. Magnetic field lines are tangled due to the Weibel instability, which allows the parallel electron heating by the shock potential. The constrained model of the electron heating predicts the ion-to-electron temperature ratio within observed values at SNR shocks and in Saturn's bow shock.

\end{abstract}

\keywords{acceleration of particles, instabilities, ISM -- supernova remnants, methods -- numerical, plasmas, shock waves}

\section{Introduction}\label{introduction}

Collisionless shocks are commonly observed in such places as planetary systems, supernova remnants (SNRs) and jets of active galactic nuclei when plasmas move with super-sonic velocities.
In such flows, collective particle interactions form a shock layer on kinetic plasma scales much smaller than the collisional mean free path. In the shock transition part of the bulk kinetic energy is converted into energies of thermal particles and electromagnetic fields through wave-particle interactions. The microphysics of these processes is still not fully understood, especially in high Mach number regime.

High Mach number shocks are supercritical, \ab{ which means that the upstream kinetic energy can not be entirely dissipated via Joule \jn{heating} (Ohmic dissipation).}
If Mach number is above \jn{a} certain limit \citep[$\ms \gtrsim 2.76$,][]{Marshall1955} part of the kinetic energy is dissipated via ion reflection by the shock potential. In quasi-perpendicular shocks the interaction of reflected ions with the upstream plasma leads to excitation of two-stream instabilities. The electrostatic Buneman instability \citep{1958PhRvL...1....8B} is excited at the leading edge of the shock foot and it results from interaction of the hot reflected ions and the cold upstream electrons. Deeper in the shock foot the interaction of reflected and upstream ions results in \jn{the} Weibel instability \citep{1959PhFl....2..337F}. These instabilities shape the shock structure forming the shock foot with electrostatic Buneman waves and Weibel filamentary structures, strongly turbulent ramp and overshoot regions, and the shock downstream filled with thermalized plasma.


This paper is the fourth in a series of works focusing on the analysis of high-Mach-number perpendicular shock physics by dint of Particle-in-Cell (PIC) simulations. 
Previously \citep[hereafter Papers I, II and III]{Bohdan2019a,Bohdan2019b,Bohdan2020} we discussed processes responsible for a production of nonthermal electrons, namely shock surfing acceleration (SSA, Paper I), its influence on the nonthermal downstream electron population (Paper II) and magnetic reconnection (Paper III). In these papers we have defined how acceleration efficiencies depend on the sonic and Alfv\'enic Mach numbers and ion-to-electron mass ratio (hereafter referred to as the mass ratio). We found that regardless of \jn{the} Mach number SSA negligibly contributes to the nonthermal electron population in perpendicular shocks and realistic mass ratio. Magnetic reconnection appears to be very active and operates as efficient electron accelerator in shocks with higher Mach numbers, however the nonthermal electron production efficiency does not depend on the mass ratio used in simulations.


Here we study electron heating processes and magnetic field amplification at high Mach number shocks with $\ma \gtrsim 20$. It was already reported that PIC simulations of low Mach number shocks \citep{2020arXiv200211132T} demonstrate good consistency of simulations and in-situ measurements of the Earth's \citep{1988JGR....9312923S} and Saturn's \citep{2011JGRA..11610107M} bow shocks. The super-adiabatic  electron heating \ab{(above the limit predicted by the Rankine-Hugoniot condition)} is associated with the cross-shock potential and interaction with ion-scale waves in the shock transition. However there is a \ab{lack of understanding how electrons are heated in the high Mach number regime.}

If  
no energy exchange between electrons and ions  is expected \jn{in the shock transition region, then}
the downstream electron temperature is described by the Rankine-Hugoniot condition.
Observations of Balmer-dominated shocks SNRs \citep{2005AdSpR..35.1017R,2008ApJ...689.1089V,2013SSRv..178..633G} and in-situ measurements of Saturn's bow shock \citep{2011JGRA..11610107M} reveal the downstream electron-to-ion temperature ratio (hereafter referred to as the temperature ratio) 
to be in \jn{a} range of $\te/\ti \approx 0.05-0.5$, which is well above the Rankine-Hugoniot prediction of $\te/\ti \approx \me/\mi$. In our previous studies \citep[][Paper II]{Bohdan2017} we reported that the electron temperature observed in PIC simulations is considerably higher \ab{than} predicted by the Rankine-Hugoniot conditions, but at the same time an energy equipartition between ions and electrons is not reached and electrons are colder than ions. Similar results were demonstrated in the PIC simulations of \cite{kato_2010},  in which $\te/\ti \approx 0.33$. These results 
suggest that super-adiabatic electron heating \jn{occurs} 
in high Mach number shocks.

\jn{We have} previously discussed that electrons can be accelerated to relativistic energies via \jn{a} number of mechanisms: SSA \citep{2000ApJ...543L..67S,2002ApJ...572..880H},  magnetic reconnection~\citep{Matsumoto2015}, \ab{stochastic Fermi-like acceleration~\citep[SFA,][]{Bohdan2017},} and stochastic shock drift acceleration~\citep{Matsumoto2017}. 
\ab{In addition \jn{to accelerating particles}, these processes cause some amount of electron heating, however the relative contribution from each is yet to be determined.} Electrons also can be heated via the shock potential which is widely discussed in low Mach number regime \citep{1987JGR....9210119T,2018PhRvL.120v5101C,2020arXiv200211132T}.

Electron heating processes are mediated by electromagnetic effects over electron temporal and spatial scales that are much shorter than the ion gyroradius or the ion skin depth. \ab{Magnetohydrodynamic (MHD)} and hybrid simulations can not describe physical processes on such small scales. 
Thus, fully kinetic simulations are needed for a proper description of the electron physics. In this paper we constrain the heating model which aims to predict the temperature ratio at high Mach number shocks. The paper is organized as follows. We present a short description of simulation setup in Section~\ref{sec:setup}. The results are presented in Section~\ref{results}. The discussion and summary are in Section~\ref{summary}.

\section{Simulation Setup} \label{sec:setup}

   \begin{table*}[!t]
      \caption{Simulation Parameters}
         \label{table-param}
     $$ 
\begin{array}{p{0.07\linewidth}ccccccccr}
\hline
\hline
\noalign{\smallskip}
Runs  & \mi/\me &   \ma & \multicolumn{2}{c}{\ms} & \multicolumn{2}{c}{\beta_{\rm e}} \\
 & & & ^*1 & ^*2 & ^*1 & ^*2 \\
\noalign{\smallskip}
\hline
\noalign{\smallskip}
A1, A2   & 50  &   22.6  & 1104 & 35  & 5 \cdot 10^{-4} & 0.5 \\
B1, B2   & 100 &   31.8  & 1550 & 49  & 5 \cdot 10^{-4} & 0.5 \\
C1, C2   & 100 &   46    & 2242 & 71  & 5 \cdot 10^{-4} & 0.5 \\
D1, D2   & 200 &   32    & 1550 & 49  & 5 \cdot 10^{-4} & 0.5 \\
E1, E2   & 200 &   44.9  & 2191 & 69  & 5 \cdot 10^{-4} & 0.5 \\
F1, F2   & 400 &   68.7  & 3353 & 106 & 5 \cdot 10^{-4} & 0.5 \\
\noalign{\smallskip}
\hline
\end{array}
     $$ 
\tablecomments{Parameters of simulation runs described in this paper. Listed are: the ion-to-electron mass ratio $\mi/\me$, and Alfv\'enic , $\ma$,  and sonic Mach numbers, $\ms$,  the electron plasma beta, $\beta_{\rm e}$.  The last two values are shown separately for the \emph{left} (runs *1) and the \emph{right} (runs *2) shock. 
All runs use the in-plane magnetic field configuration $\varphi=0^o$.} 
   \end{table*}

Simulations are performed \jn{with} 
the modified version of the relativistic electromagnetic TRISTAN code \citep{Buneman1993} with  MPI-based parallelization \citep{2008ApJ...684.1174N,2016ApJ...820...62W} and  the particle sorting optimization \citep{10.1007/978-3-319-78024-5_15}. The Vay solver \citep{Vay2008} is used to update particle positions. 
The triangular-shape-cloud particle shapes (the second-order approximation) and \cite{Friedman1990} filter for electric and magnetic fields are used to suppress the numerical grid-Cherenkov short-wave radiation. To perform shock simulations we use a 2D3V code configuration which follows two spatial coordinates and all three components of \jn{particle} velocities and 
electromagnetic fields.


The flow-flow simulation setup is used to initialize shocks. Note that the same setup was used in our previous works (\citet{Bohdan2017}, Papers I, II and III) and more detailed description can be found in Paper I. The simulation setup assumes collision of two counterstreaming electron-ion plasma flows which leads to the formation of two shocks separated by a contact discontinuity.  Here we refer to shocks as the \emph{left} and the \emph{right} shocks. The absolute values of the beam velocities are equal, $v_{\rm L}=v_{\rm R}=v_{\rm 0}=0.2c$.
Plasma beams are  equal in density but 
\jn{their temperatures}
differ by \jn{a} factor of $1000$. Thus the \emph{electron} plasma beta (the ratio of the electron plasma pressure to the magnetic pressure) for the left beam is $\beta_{\rm e,L}=5 \cdot 10^{-4}$ and $\beta_{\rm e,R}=0.5$ for the right beam.

The large scale upstream magnetic field makes \jn{an} angle $\varphi=0^o$ with the simulation plane \jn{(the} so-called \emph{in-plane} configuration). Note, that such simulations give us a good representation of the 3D shock physics \citep{Bohdan2017,Matsumoto2017}. The adiabatic index is thus $\Gamma=5/3$.  The resulting shock speed equals $\vsh=0.263c$ in the \emph{upstream} reference frame. 
The Alfv\'en velocity is defined as $v_{\rm A}=B_{\rm 0}/\sqrt{\mu_{\rm 0}(N_e\me+N_i\mi)}$, where $\mu_{\rm 0}$ is the vacuum permeability, $N_i$ and $N_e$ are the ion and the electron number densities, and $B_0$ is the far-upstream magnetic-field strength. The sound speed reads $c_{\rm s}=(\Gamma k_BT_{\rm i}/\mi)^{1/2}$, where $k_B$ is the Boltzmann constant and $T_{\rm i}$ is the ion temperature defined as $k_BT_i=m_i v_{th}^2/2$, \jn{where} $v_{th}$ is defined as the most probable speed of the upstream plasma particles in the upstream reference frame.
The Alfv\'enic, $\ma=\vsh/v_{\rm A}$, and sonic, $\ms=\vsh/c_{\rm s}$, Mach numbers of the shocks are defined in the conventional \emph{upstream} reference frame.
Note that the \emph{sonic} Mach number, $M_{\rm s}$, of the two shocks differ by a factor of $\sqrt{1000} \simeq 30$ because of \jn{the difference in} $\beta_{\rm e}$.

The ratio of the electron plasma frequency, $\omega_{\rm pe}=\sqrt{e^2N_e/\epsilon_0\me}$, to the electron gyrofrequency, $\Omega_{\rm e}=eB_0/\me$, \jn{is} in the range \jn{of} $\omega_{\rm pe}/\Omega_{\rm e}=8.5-17.3$. Here, $e$ is the electron charge, and $\epsilon_0$ is the vacuum permittivity.
The electron skin depth in the upstream plasma is common for all runs and equals $\lse=20\Delta$, where $\Delta$ is the size of grid cells. As the unit of length we use the ion skin depth, $\lsi=\sqrt{\mi/\me}\lse$. As the time-step $\delta t=1/40\,\omega_{\rm pe}^{-1}$ is used. 
The time scales are 
\jn{given} in terms of the upstream ion Larmor frequency, $\Omega_{\rm i}$, where $\Omega_{\rm i}=eB_0/\mi$. The number density in the far-upstream region is 20 particles-per-cell for each species.

In the following sections, we discuss the results of six  large-scale numerical experiments (runs A--F), featuring in total twelve \pjm{physically distinct} simulated shocks. We therefore refer to each of these shock cases as to a separate simulation run, and label the shocks in the left plasma (cold shock, $\beta_{\rm e,L}=5 \cdot 10^{-4}$) with *1, and the right shocks with *2 (warm shock, $\beta_{\rm e,R}=0.5$).
The derived parameters of the simulation runs described in this paper are listed in Table~\ref{table-param}. If the result is valid for both cold and warm shocks we use only a letter without a digit.

These simulations cover a wide range of mass ratios ($\mi/\me = 50-400$) and Alfv\'enic Mach numbers ($\ma=22.6-68.7$), thus permitting a \pjm{thorough} investigation of the influence of these parameters on the electron heating processes. The goal of this paper is to clarify these dependencies in order to give a prediction for realistic high Mach number shocks.

\section{Results} \label{results}

   \begin{table}[!t]
      \caption{Thermal properties of the downstream plasma}
         \label{table-temp}
\centering
\begin{tabular}{lcccccccc}
\hline
\hline
\noalign{\smallskip}
Run &  $\dfrac{k_{\rm B}\te}{\me c^2}$ & $\dfrac{k_{\rm B}\te}{\me c^2}$  &  $\te/\ti$ & $\te/\ti$  \\

 & (Paper II) & (eq.~\ref{total-heating})  &   & (eq.~\ref{temp-ratio})  \\

\noalign{\smallskip}
\hline
\noalign{\smallskip}
\multirow{1}{*}{A1} & 
\multirow{1}{*}{$0.107\pm 0.004$} & 
\multirow{2}{*}{$0.1\pm 0.03$} &
\multirow{1}{*}{$0.17\pm 0.04$}  & 
\multirow{2}{*}{$0.16\pm 0.05$} \\
\multirow{1}{*}{A2} & 
\multirow{1}{*}{$0.091\pm 0.004$} &   & \multirow{1}{*}{$0.14\pm 0.03$} &  \\
\noalign{\smallskip}
\hline
\noalign{\smallskip}
\multirow{1}{*}{B1} & 
\multirow{1}{*}{$0.216\pm 0.004$} & 
\multirow{2}{*}{$0.19\pm 0.05$} &
\multirow{1}{*}{$0.18\pm 0.03$}  & 
\multirow{2}{*}{$0.16\pm 0.05$} \\
\multirow{1}{*}{B2} & 
\multirow{1}{*}{$0.183\pm 0.004$} &   & \multirow{1}{*}{$0.15\pm 0.03$} &  \\
\multirow{1}{*}{C1} & 
\multirow{1}{*}{$0.253\pm 0.002$} & 
\multirow{2}{*}{$0.22\pm 0.07$} &
\multirow{1}{*}{$0.19\pm 0.03$}  & 
\multirow{2}{*}{$0.19\pm 0.06$} \\
\multirow{1}{*}{C2} & 
\multirow{1}{*}{$0.217\pm 0.003$} &   & \multirow{1}{*}{$0.17\pm 0.02$} &  \\
\noalign{\smallskip}
\hline
\noalign{\smallskip}
\multirow{1}{*}{D1} & 
\multirow{1}{*}{$0.332\pm 0.024$} & 
\multirow{2}{*}{$0.34\pm 0.08$} &
\multirow{1}{*}{$0.15\pm 0.03$}  & 
\multirow{2}{*}{$0.15\pm 0.04$} \\
\multirow{1}{*}{D2} & 
\multirow{1}{*}{$0.28\pm 0.003$} &   & \multirow{1}{*}{$0.11\pm 0.01$} &  \\
\multirow{1}{*}{E1} & 
\multirow{1}{*}{$0.394\pm 0.005$} & 
\multirow{2}{*}{$0.38\pm 0.1$} &
\multirow{1}{*}{$0.16\pm 0.02$}  & 
\multirow{2}{*}{$0.17\pm 0.05$} \\
\multirow{1}{*}{E2} & 
\multirow{1}{*}{$0.368\pm 0.009$} &   & \multirow{1}{*}{$0.16\pm 0.02$} &  \\
\noalign{\smallskip}
\hline
\noalign{\smallskip}
\multirow{1}{*}{F1} & 
\multirow{1}{*}{$0.765\pm 0.035$} & 
\multirow{2}{*}{$0.8\pm 0.2$} &
\multirow{1}{*}{$0.18\pm 0.04$}  & 
\multirow{2}{*}{$0.18\pm 0.06$} \\
\multirow{1}{*}{F2} & 
\multirow{1}{*}{$0.732\pm 0.02$} &   & \multirow{1}{*}{$0.18\pm 0.04$} &  \\
\noalign{\smallskip}
\hline
\end{tabular}
\tablecomments{Electron temperature and electron-to-ion temperature ratio in the shock downstream. Second column - the downstream electron temperatures listed in Paper II. Third column - the downstream electron temperature calculated via equation~\ref{total-heating}. Fourth column - the electron-to-ion temperature ratio observed in simulations. Fifth column - the electron-to-ion temperature ratio calculated via equation~\ref{temp-ratio}.  } 
   \end{table}

\begin{figure*}[htb]
\centering
\includegraphics[width=0.99\linewidth]{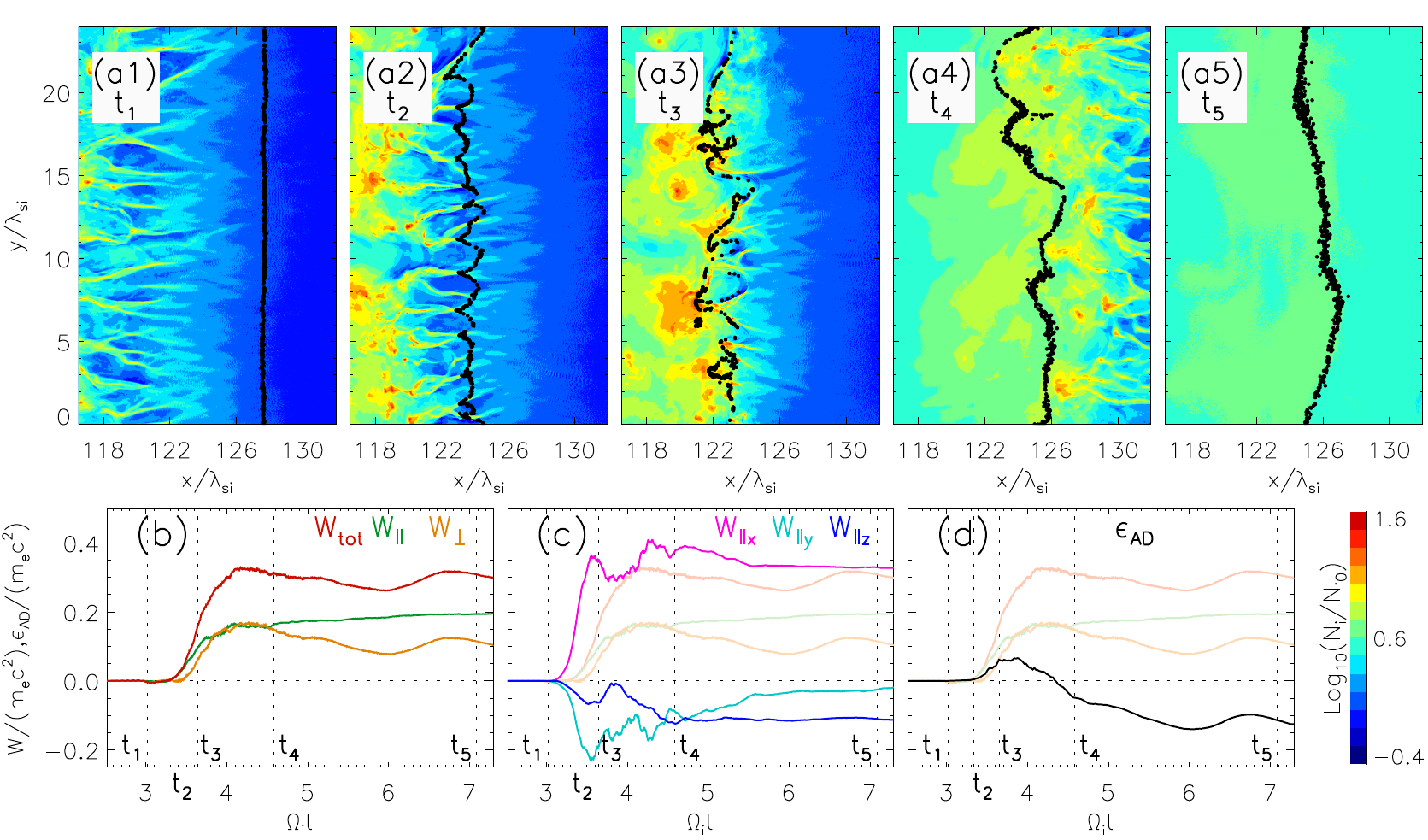}
\caption{Evolution of the work components of traced electrons for run B2. Panels (a1)–(a5) present the positions of traced electrons (black dots) at the time t1–t5 on density maps of the shock region.  
Panel (b): the temporal evolution of the total work done by electric field ($\wtot$, red line), the parallel work component ($W_\parallel$, green line) and the perpendicular work component ($W_\perp$, orange line).
\ab{Panel (c): magenta, light blue and dark blue lines are x- ($W_{\parallel x}$), y- ($W_{\parallel y}$) and z-components ($W_{\parallel z}$) of the parallel work.}
Panel (d): evolution of the adiabatic component \ab{($W_{AD}$, see Section~\ref{sec:adiabat})}. Faded curves in panel (c) and (d) represent lines from panel (a). The vertical dashed lines in panels (b), (c) and (d) are time markers for $t_1-t_5$, the horizontal dashed line is the zero level.}
\label{ele-heating}
\end{figure*}

In Paper II we showed that the downstream electron temperature is approximately proportional to the mass ratio, therefore a similar fraction of the upstream ion kinetic energy is transferred to thermal electrons in the simulations discussed here. It also results in similar temperature ratios in all runs (see Table~\ref{table-temp}).


Here we want to constrain the electron heating model based on simulation data. First, we identify processes responsible for the electron heating. Next, using known properties of mechanisms and particle data, we determine how individual heating efficiencies depend on the mass ratio ($\mi/\me$), Alfv\'enic Mach number ($\ma$) and the shock velocity ($\vsh$). Finally, using energy gains from individual heating processes, we calculate numerical coefficients for them in \ab{order to confirm} the heating model and simulation results.


\subsection{Electron energy evolution}

Particle energies change due to work, $\wtot$, done by the electric field. 
For identification of heating process we split this work into two parts, namely, the work done in the directions parallel, $W_\parallel$, and perpendicular, $W_\perp$, to the local magnetic field. Therefore the electron kinetic energy is calculated as 
\be
\varepsilon = \wtot+ \varepsilon_0 = W_{\parallel} + W_{\perp} +  \varepsilon_0 \ ,
\ee
where $\varepsilon_0$ is an initial kinetic energy,  $W_{\parallel}=\int_{t'=0}^t e E_{\parallel} v_{\parallel} dt'$ and $W_{\perp}=\int_{t'=0}^t e E_{\perp} v_{\perp} dt'$. 
All variables denoted with 
$\perp$ \jn{or} $\parallel$ are vector components perpendicular and parallel to the local magnetic field, \jn{respectively}.
Taking into account that the electron downstream temperature is approximately proportional to the mass ratio, we derive energy incomes of individual processes in terms of the ion upstream kinetic energy, $\mi\vsh^2$. Note, that energy incomes strongly vary during shock self-reformation cycles, thus in our calculation we use values averaged over one reformation cycle.

To demonstrate the main stages of the electron heating we use the energy history of electrons selected in the shock upstream of run B2 (Fig.~\ref{ele-heating}). Electrons reside at the leading edge of the shock foot with electrostatic Buneman waves at $t_1$ (panel (a1)), the Weibel instability region of the foot at $t_2$ (panel (a2)), the ramp \jn{region} where magnetic reconnection occurs at $t_3$ (panel (a3)), the turbulent overshoot at $t_4$ (panel (a4)) and the downstream at $t_5$ (panel (a5)).

\ab{The total energy change of the electron population is represented} by $\wtot$ in Figure~\ref{ele-heating}(b). The energy of the electrons starts to grow when they enter the shock foot at $t_2$ (panel (a2)). Figure~\ref{ele-heating}(b) shows that the energy reaches the maximal value at $\omci t \approx 3.7$ when particles reside in the shock overshoot, and then it gradually decreases when electrons are advected to the shock downstream. At the final step about 70\% of the total energy is produced by parallel heating and the remaining part has a perpendicular origin.

We have identified five processes responsible for the electron heating: SSA via interaction with Buneman waves at the leading edge of the shock foot, a heating by the shock potential at the shock foot and the ramp, magnetic reconnection in the shock ramp, \ab{SFA in the shock} ramp and the overshoot, and an adiabatic heating which follow plasma compression and decompression.

\subsection{Heating via SSA}

\pjm{The} \ab{Buneman instability is excited because of} \pjm{the} \ab{relative motion of the upstream electrons} \pjm{to the} \ab{reflected ions,} \pjm{with the energy budget available from this process approximately given by $E_{\rm SSA,avail} \approx 2 \me \vsh^2 \simeq 0.15\me c^2$} \citep{2009ApJ...690..244A}. \ab{Later this energy is converted into electron heating, therefore it can be used as} \pjm{an upper-limit for the energy provided to electrons via SSA}.
In Paper II we discussed that only about 1\% of the available energy is transferred to the \emph{nonthermal} high energy electrons. However, in this case we are interested in the \emph{total} energy gain, which is about $0.1 E_{\rm SSA,avail}$ for the selected population of electrons.
Thus we estimate the energy gain via SSA as 

\be
\varepsilon_{\rm SSA} \approx \alpha_{\rm SSA} \me \vsh^2 \ = \alpha_{\rm SSA} \frac{\me}{\mi} \mi \vsh^2 \ ,
\label{heat_SSA}
\ee
where $\alpha_{\rm SSA}$ depends on the magnetic field configuration of the simulation and it equals $\alpha_{\rm SSA,0}=0.2 \pm 0.1$ for the in-plane case \pjm{, where the subscript `0' denotes the angle of the initial magnetic field to the simulation plane}. For \jn{the} selected electron set the heating via SSA \jn{takes place} at $\omci t \approx 3$, it is not well visible in Figure~\ref{ele-heating} because of the very small efficiency compared to other processes. Note that SSA contributes to the perpendicular work component. It is also easy to see \jn{from equation~\ref{heat_SSA}} that the heating due to SSA may contribute a substantial  amount of energy only in case of small mass ratios. 

Previously we demonstrated \jn{in} \citet{Bohdan2017,Matsumoto2017} that the realistic  3D SSA efficiency is reproduced in simulations with an out-of-plane \jn{magnetic field} configuration ($\varphi=90^o$) where Buneman waves are well captured in the simulation plane. In runs with such 
\jn{a field} 
configuration, the heating efficiency is higher, with around 50\% of the available energy transferred and $\alpha_{\rm SSA,90}=1 \pm 0.5$. This value \jn{we put} in the heating model of realistic shocks.

\subsection{Heating by the shock potential}

In our simulations electrons are magnetized, thus they follow its initial magnetic flux tube and can freely move only along magnetic field lines. Therefore if any plasma disturbance  creates regions with aligned magnetic and electric fields (which initially are perpendicular to each other) electrons can be energized via parallel heating. 
At \jn{a} shock transition of high Mach number shocks \jn{the} Weibel instability and magnetic reconnection are responsible for generation of turbulent magnetic fields

\begin{figure}[htb]
\centering
\includegraphics[width=0.70\linewidth]{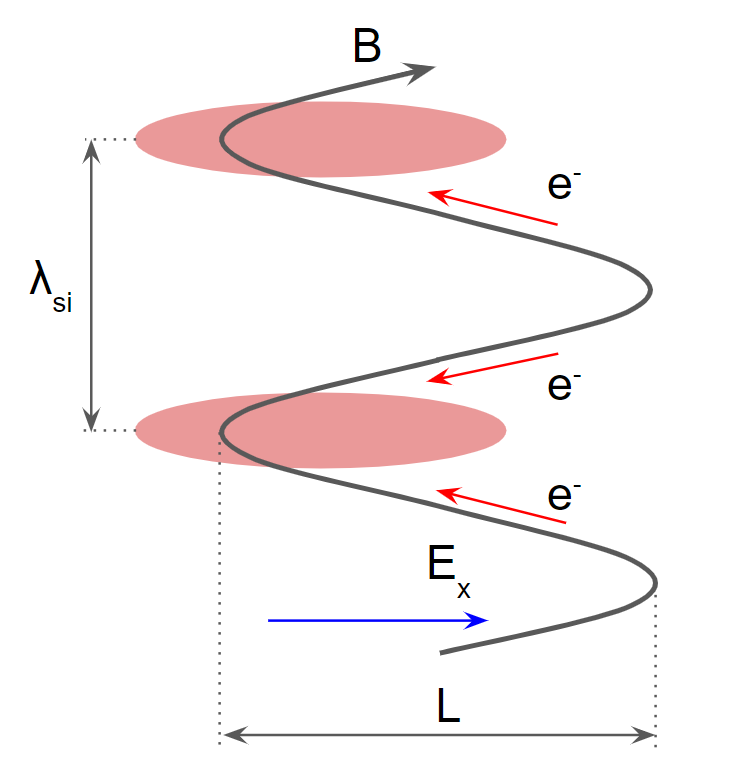}
\caption{Scheme of electron heating by the shock potential in the shock foot. Red ellipses indicate dense Weibel filaments. Electron motion is designated by red arrows. Blue arrow is the shock potential, $E_x$, in the shock foot. $\lsi$ represents the Weibel instability wavelength and $L$ is a measure of a magnetic field line widening because of $B_x$ amplification.}
\label{ele-heating-SP}
\end{figure}

In Figure~\ref{ele-heating}(a2), the electron positions show that magnetic field lines are deformed because of additional $B_x$ and $B_z$ \jn{components} generated by the Weibel instability. In addition to that, the shock potential ($E_x$) starts to operate in this region. The magnetic field structure is represented by non-propagating linearly polarized magnetic field waves generated by the Weibel instability (Fig.~\ref{ele-heating-SP}). This field is regular with the spatial scale of $\lsi$. Deeper in the shock, compressions and magnetic reconnection make the magnetic field structure more turbulent and chaotic (Fig.~\ref{ele-heating}(a3)). 
Therefore an initial upstream field configuration is destroyed in the shock transition, creating suitable conditions for the parallel heating of electrons.

Parallel acceleration for selected portion of electrons occurs in two stages. During the first stage, $\omci t = 3.1-3.6$, electrons reside \jn{in} the shock foot with regular filamentary magnetic field produced by \jn{the} Weibel instability. The electron energy increases because of a rapid growth of $W_\parallel$. During the second stage, $\omci t = 3.6-5$, electrons are at the shock ramp-overshoot, where the magnetic field is turbulent, and  $W_\parallel$ grows slower than in the first stage. 
Here we call these parts of the parallel work as the regular component, $\varepsilon_{\rm SP,r}$, and the turbulent component, $\varepsilon_{\rm SP,t}$, because of the magnetic field structure at the corresponding regions. Note that 'SP' refers to the shock potential.

While the Weibel instability develops, electrons are moving to the left and forming dense filaments (Fig.~\ref{ele-heating-SP}).  During this motion they gain some energy because of the work done by $E_x$-field (see $W_{\parallel,x}$ in Figure~\ref{ele-heating}(c)). We found that in this region $W_{\parallel,x}$ is always the main energy contributor and for further discussions we ignore $W_{\parallel,y}$ and $W_{\parallel,z}$. Thus the energy income is
\be
\begin{split}
\varepsilon_{\rm SP,r} \propto eE_x L = eE_x \frac{B_x}{B_y} \frac{\lsi}{2} & \propto e E_x \ma \frac{\lsi}{2} \propto \\
& \propto \frac{1}{2} \mi \vsh^2 \ .
\end{split}
\label{heat_SP0} 
\ee
Here we use two relations derived from simulation data. \ab{The saturation level of the normalized ${B_x}$ generated by \jn{the} Weibel instability is roughly proportional to the Alfv\'enic Mach number, so the ratio of $\dfrac{1}{\ma} \dfrac{B_x}{B_y}$ is constant.} 
\ab{We also can express $E_x$ in terms of the upstream fields.} \ab{The shock potential, $e \phi$, is proportional to the ion upstream kinetic energy. We can \pjm{therefore} estimate the electric field as $E_x \propto \phi / L_{\rm ramp} \propto r_{\rm g} = \vsh/\omci$, then $E_x = \eta \vsh B_0$. In Paper III we discussed that all shocks host very similar conditions at the shock ramp, namely the same fraction of reflected ions, the same relative velocities etc. Therefore the coefficient $\eta$ appears to be the same in our simulations.}

Using the observed parallel energy income during this stage, we can derive the regular heating component:
\be
\varepsilon_{\rm SP,r}= \alpha_{\rm SP,r} \mi \vsh^2 \ ,
\label{heat_SP1}
\ee
where $\alpha_{\rm SP,r}=0.012 \pm 0.002$.

The ramp-overshoot region is strongly turbulent thus the scheme presented in Figure~\ref{ele-heating-SP} is no longer valid. However the thorough investigation of $W_{\parallel}$ at this region demonstrates that the main energy contributor is still the $x$-component of the electric field.
Note that a part of $W_{\parallel}$ during the second stage is explained by magnetic reconnection (see section~\ref{sec:MR}) but we account for it in our calculations.

As soon as the magnetic field is turbulent electrons have a possibility to move some distance along the $x$-axis, which is proportional to the length of the turbulent region, $L \approx L_{\rm ramp} \propto r_{\rm gi} = \ma \lsi$. As we discussed above the average electric field is proportional to the upstream electric field, $E'_x \propto \vsh B_0$. The only difference \jn{is} that this field is opposite (negative) to the shock potential which heats electrons during the first stage.
Therefore the chaotic component is also proportional to the upstream ion kinetic energy, $\varepsilon_{\rm SP,t} \propto \mi \vsh^2$. Using the simulation data we can write that
\be
\varepsilon_{\rm SP,t}= \alpha_{\rm SP,t} \mi \vsh^2 \ ,
\label{heat_SP2}
\ee
where $\alpha_{\rm SP,t}=0.013 \pm 0.002$ which \ab{is consistent (as error bars overlap) with what} we estimated for the regular component. Such a result is not surprising because during this process magnetic field lines are straightening and electrons are redistributed back evenly along magnetic field lines in the presence of a negatively directed $E_x$. It is essentially the opposite to what happens in the Weibel instability region, \jn{albeit} in more turbulent and chaotic way.

\subsection{Heating by magnetic reconnection}\label{sec:MR}

In Paper III we discussed how magnetic reconnection accelerates electrons to nonthermal energies. We also found that magnetic reconnection becomes more active in shocks with high Alfv\'en Mach numbers and the fraction of electrons involved in magnetic reconnection correlates with $\ma$. It makes the heating via magnetic reconnection strongly dependent on Alfv\'en Mach number.

For the selected electron population magnetic reconnection takes place at $\omci t \approx 3.5-3.6$. \ab{In Paper II we discussed that magnetic reconnection can contribute both to $W_\perp$ (e.g., Fermi-like processes) and $W_\parallel$ (e.g., parallel heating at x-points).}
For run B2 the average energy of the electrons that have undergone reconnection is $0.21\me c^2$. 
Heating via SSA and the regular part of the shock potential occurs before magnetic reconnection, thus energy gained by electrons due to magnetic reconnection is $\varepsilon_{\rm MR,sim} = 0.21\me c^2 - \varepsilon_{\rm SSA} - \varepsilon_{\rm SP,r} = 0.115\me c^2$. Taking into account that about \ab{26\% (see Paper III)} of electrons are involved in magnetic reconnection, the fraction of the total heating delivered by magnetic reconnection is about 11\% in run B2. 

The number of vortices generated during magnetic reconnection  differs by a factor of 60 between runs A and F (see Table 2. in Paper III). However the heating efficiency grows not so quickly. Magnetic reconnection gives 6\% of the total heating in runs A, 11\% in runs B, 18\% in runs C, 15\% in runs D, 17\% in runs E, 20\% in runs F. It happens because the electron temperature at reconnection sites, $T_{\rm MR}$, becomes smaller compared to the downstream electron temperature, $T_{e}$, in high $\ma$ cases. $T_{\rm MR}/T_e$ is about 1 in runs A and 0.5 in runs F.

To approximate the heating via magnetic reconnection we also need to determine boundary conditions. We assume that magnetic reconnection is switched-off when $\ma < 20$, in Paper III we have already demonstrated that in run A the number of reconnection sites is very small and it drops quickly.
In the high Mach number limit ($\ma \gtrsim 100$) we expect saturation of magnetic reconnection efficiency at a level which does not exceed 25\%, because already in run F almost all electrons are involved in the process and the ratio $T_{\rm MR}/T_e$ will likely stabilize around 0.5. Therefore the heating due to magnetic reconnection we estimate as
\be
\begin{split}
\varepsilon_{\rm MR} & = \alpha_{\rm MR} \mi \vsh^2 \ , \\
\alpha_{\rm MR} & = \left\{
  \begin{array}{ll}
    0   & , \ma < 20\\
    \alpha^*_{\rm MR} (\ma-20)^\frac{1}{3} \ & , 20 <\ma < 100\\
    \alpha^*_{\rm MR} \ 80^\frac{1}{3} & , \ma > 100
  \end{array}
\right.
\end{split}
\label{heat_MR}
\ee
where $\alpha^*_{\rm MR}=(22\pm 4)\cdot 10^{-4}$. For the discussion of the heating model in Section~\ref{heat-model} we assume that magnetic reconnection contributes equally to $W_\perp$ and $W_\parallel$.

\subsection{Heating via SFA}

In \cite{Bohdan2017} we have discussed that high-energy electrons are  accelerated via SFA. However this process works regardless of the particle energy and also contribute to the electron heating.

During the time interval of $\omci t \approx 3.6-4.5$ electrons are in the shock ramp-overshoot region (Fig.~\ref{ele-heating}(a3),(a4)) and the perpendicular work component grows (Fig.~\ref{ele-heating}(b)). This region is characterized by a turbulent magnetic field hosting appropriate conditions for SFA. When a particle interacts with moving magnetic field structures it is accelerated by a motional electric field, $\mathbf{E}=-\mathbf{v}\times\mathbf{B}$, thus SFA contributes to $W_{\perp}$.

To estimate energy income in SFA we need to know a number of collisions of electron with scattering centers, $N_{coll}$, and the average velocity of these centers, $v_{mag}$. 
The energy income via SFA is 
\be
\begin{split}
\frac{\varepsilon_{\rm SFA}}{\varepsilon_e} & \propto \left( \frac{v_{\rm mag}}{v_e}\right)^2 N_{\rm coll}  = \left( \frac{v_{\rm mag}}{v_e}\right)^2 \frac{T}{\Delta t} \propto 
\\ 
 & \propto \left( \frac{v_{\rm mag}}{v_e}\right)^2 \frac{r_{\rm gi} v_e}{\vsh \lsi} = \ma \frac{v_{\rm mag}^2}{v_e \vsh} \ ,
\end{split}
\ee
where $\varepsilon_e$ and $v_e$ is the electron energy and velocity before SFA, $N_{\rm coll}= T/\Delta t$, where $T \propto r_{\rm gi}/\vsh$ is the total time an electron spends in the shock transition, $\Delta t \propto \lsi/v_e$ is the average time between collisions, where the average distance between scattering centers are defined by Weibel instability with the spatial scale of $\lsi$.
In our simulations we found that $v_{\rm mag}$ does not depend on the mass ratio or Mach number and it can be represented as $v_{\rm mag}= \chi \vsh $, where $\chi\approx0.25$ was derived from simulations.

Heating by SFA occurs after energization by SSA, the regular part of the shock potential and magnetic reconnection, thus we can assume that initial energy for SFA is  $\varepsilon_e=(\me v_e^2)/2=\varepsilon_{\rm SSA}+\varepsilon_{\rm SP,r} +\varepsilon_{\rm MR}$. Therefore the energy income from SFA is:
\be
\begin{split}
\varepsilon_{\rm SFA} & = \alpha_{\rm SFA} \mi \vsh^2 \ , \\
\alpha_{\rm SFA} & = \alpha^*_{\rm SFA} \ma \chi^2 \sqrt{\frac{(\alpha_{\rm SSA} + \alpha_{\rm SP,r}+\alpha_{\rm MR}) \me}{2\mi}}
\end{split}
\label{heat_SFA}
\ee
where $\alpha^*_{\rm SFA}=0.7\pm 0.2$.

Here we used nonrelativistic formulas because it has a minor effect for cases discussed here and it fits well for nonrelativistic SNR shocks or planetary bow shocks.

\subsection{Adiabatic heating}\label{sec:adiabat}

In the shock transition plasma is compressed, thus we need to account for a heating because of adiabatic compression. During compression the plasma temperature can be approximated by
\be
T V^{\Gamma-1}=const
\label{eq:8}
\ee
where $V$ is the volume of the flux tube and $\Gamma=5/3$ is the adiabatic index. Using the total derivative of equation~\ref{eq:8} we estimate the energy income from adiabatic compression as
\be
\begin{split}
\varepsilon_{\rm AD} & =\sum_{n} \varepsilon_{\mathrm{AD},n} \ ,  \\
\varepsilon_{\mathrm{AD},n} & = k_{\rm B} T_n \left(1-\frac{V_{n+1}}{V_{n}}\right) \ .
\end{split}
\label{heat_AD1}
\ee
and sum is done over time steps while electrons travel from the upstream to the downstream. Here we use B-field maps to derive the volume of the flux tube and the traced particle data to estimate electron temperature inside the selected flux tube. 

Overall impact of a compression-decompression cycle appears to be negative (Fig.~\ref{ele-heating}(d)). Such result is slightly counterintuitive, however it can happen if an additional heating process operates during a compression-decompression cycle. 
\ab{Initially the electron temperature is low thus the adiabatic heating during compression is also low. However decompression happens when electrons cross the shock overshoot and they are already hot. Therefore the adiabatic cooling is high and in our case it overcomes the adiabatic heating during the compression. It results in overall negative impact of the compression-decompression cycle.}

As expected the behaviour of the adiabatic energy income is consistent with $W_\perp$ and $\wtot$ in the shock downstream ($\omci t > 5$) where only adiabatic compression and decompression is responsible for the electron energy change. 

Compression operates on top of all discussed above processes and adiabatic influence can be represented as a modulation of energy incomes from processes contributing to $W_\perp$. Thus
\be
\begin{split}
W_\perp & = \alpha_{\rm AD} \varepsilon_\perp \\
\alpha_{\rm AD} & = 0.6 \pm 0.1
\end{split}
\label{heat_AD}
\ee
where  $\varepsilon_\perp$ is the sum of energy incomes from processes contributing to $W_\perp$.

\subsection{Model of the electron heating.}\label{heat-model}

To compile the heating model we combine all individual heating efficiencies (eq.~\ref{heat_SSA}, \ref{heat_SP1}, \ref{heat_SP2}, \ref{heat_MR}, \ref{heat_SFA}, \ref{heat_AD}).  Recall that both shock potential components contribute to the parallel work, magnetic reconnection contribute evenly to the parallel and perpendicular parts, and SSA, SFA and adiabatic heating affect on the perpendicular component. Therefore expanding equation~\ref{heat_AD} we can write
\be
\begin{split}
\wtot &  = \varepsilon_{\rm SP,r} +\varepsilon_{\rm SP,t} +  \varepsilon_{\rm MR}/2 + \\ 
& + \alpha_{\rm AD} (\varepsilon_{\rm SSA} +\varepsilon_{\rm SFA} + \varepsilon_{\rm MR}/2) \ ,
\end{split}
\label{total-heating}
\ee
and the downstream electron temperature reads
\be
k_{\rm B} T_e = \frac{2}{3} \left(\wtot + \varepsilon_0 \right) \ .
\label{total-temp}
\ee

The amount of available energy for the particle heating is defined by Rankine-Hugoniot relations, namely by the ion temperature jump condition. Here we neglect electron kinetic energy and the ion thermal energy in the shock upstream. The available energy is
\be
E_{\rm avail} = \frac{3}{2} T_{i} =  \frac{3}{2} \cdot \frac{5}{16} \ms^2 T_{up,i} = \frac{1}{2} \mi v_0^2.
\label{avail-ene}
\ee
Thus in the downstream reference frame, where our simulations are performed, the upstream ion kinetic energy is the source for the downstream thermal energies of ions and electrons. Therefore the temperature ratio can be calculated as
\be
\frac{\te}{\ti}= \frac{\wtot}{E_{\rm avail}-\wtot} \ .
\label{temp-ratio}
\ee
Note that here we also neglect energies of electric and magnetic fields which are small compared to other components.

The modeled downstream temperatures and temperature ratios are listed in Table~\ref{table-temp} and prediction of the heating model is consistent with the simulations results.

\section{Summary and discussion} \label{summary}

This work is the fourth paper in a series investigating physics of  nonrelativistic perpendicular high Mach number shocks. In our previous studies we discussed production of nonthermal electrons via SSA operating at the shock foot (Papers I and II) and via magnetic reconnection that results from the
nonlinear decay of ion Weibel filaments at the shock ramp (Paper III). \ab{In this paper we investigate super-adiabatic electron heating. We constrain the heating model for the downstream electron-to-ion temperature ratio in realistic high Mach number shocks. 2D simulations with an in-plane magnetic field configuration} \pjm{are capable of reproducing the main shock structures found in 3D simulations}~\citep{Bohdan2017,Matsumoto2017} \ab{and therefore electron heating is also well captured} \pjm{at a fraction of the computational expense}. \ab{Our simulations are  performed for a wide range of physical parameters ($\mi/\me=50-400$ and $\ma=22.6-68.7$) which permits predictions for realistic shocks.}

\ab{In Paper II we discussed that all shocks demonstrate super-adiabatic electron heating, and that the downstream electron temperature is well above that predicted by the Rankine-Hugoniot condition. Here we calculate the downstream temperature ratio which is in range of $\te/\ti = 0.11-0.19$.}

\ab{Electrons are heated via SSA, parallel acceleration due to the shock potential, magnetic reconnection, and SFA. \mpo{Electrons energized} via SSA, magnetic reconnection and SFA have been already extensively studied, \mpo{and} here heating via the shock potential at the shock foot is discussed for the first time. At perpendicular shocks the upstream magnetic field and the shock potential are perpendicular to each other. Nevertheless, operating at the shock foot, the Weibel instability \mpo{generates a $B_x$ magnetic field} \pjm{component} \ab{which permits parallel heating} by the shock potential. It consist of a regular part which deals with ordered Weibel filaments at the shock foot and a turbulent part which operates deeper in the turbulent ramp-overshoot region. }

\ab{In addition to these processes, compression at the shock transition influences the plasma temperature, \mpo{but its} total effect is surprisingly negative. \mpo{During passage from the upstream to the downstream  plasma is compressed to the overshoot density ($N_{max}/N_0 \approx 8$) and then relaxes and reaches a compression ratio of 4 expected for high Mach numbers. The adiabatic heating at the compression stage is smaller than cooling after the shock overshoot, on account of nonadiabatic heating in-between. Hence} the net adiabatic cooling.}

\ab{Among these processes only SSA is limited by the \emph{electron} upstream kinetic energy as} \pjm{the instability} \mpo{is driven by inbound electrons.} 
\ab{All other processes are the \mpo{ion-related} phenomena, and the incoming energy is proportional to the \emph{ion} upstream kinetic energy. They are actually responsible for the super-adiabatic electron heating. Combining all detected heating processes we constrain the heating model. It accurately predicts the downstream electron temperature of perpendicular nonrelativistic high Mach number shocks performed by means of PIC simulations.}

\begin{figure}[htb]
\centering
\includegraphics[width=0.99\linewidth]{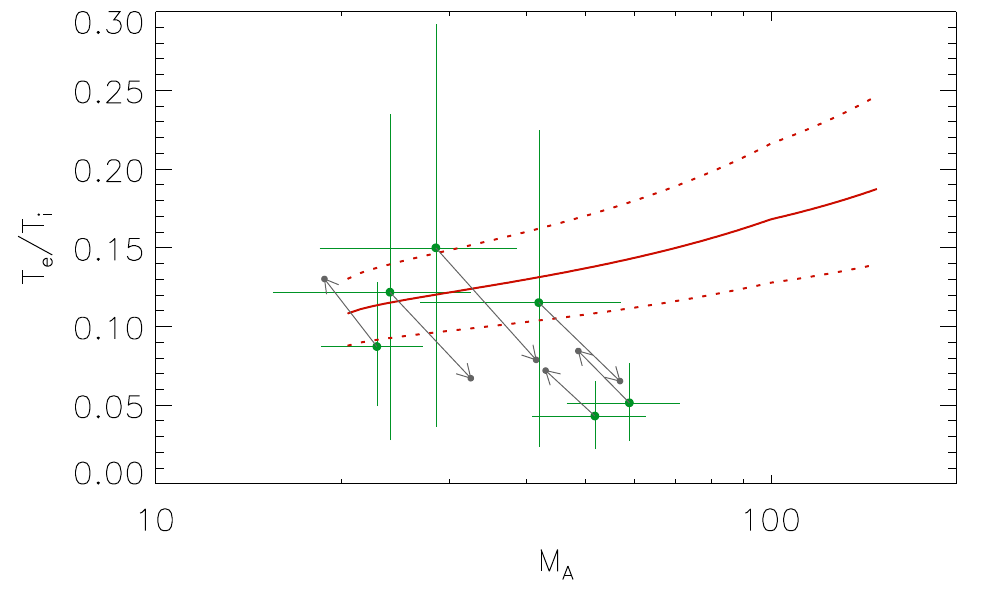}
\caption{Comparison of the model temperature ratio with Cassini measurements during Saturn's bow shock crossings with $\ma > 20$ and $\theta_{B_n} > 45^o$ \citep{2011JGRA..11610107M}. Red solid line represents the heating model (eq.~\ref{temp-ratio}) and red dotted lines are errorbars. Green dots are Cassini measurements, \mpo{with grey arrows pointing to the uncorrected values}.}
\label{total-heating-saturn}
\end{figure}

\ab{We apply the heating model to realistic shocks using the real proton-to-electron mass ratio and assuming that heating processes are not significantly different between our 2D in-plane shock simulations and real 3D shocks. 
In SNR shocks with propagation velocity above 1000 $\rm{km}/\rm{s}$, which are more than likely to be high Mach number shocks, the observed temperature ratio is $\te/\ti=0.05-0.2$ \citep{2005AdSpR..35.1017R,2008ApJ...689.1089V,2013SSRv..178..633G}, \mpo{based on X-ray spectra for $\te$ and Balmer-line width for $\ti$}. We \mpo{do not precisely know the Alfv\'enic Mach number of SNR shocks, and so we simply assume that it is} above 20. The predicted temperature ratio is $\te/\ti=0.09-0.25$ for $\ma > 20$, which indicates a good match between observations and PIC simulations.}

\mpo{At Alfv\'enic Mach numbers $\ma > 100$ the Weibel instability is heavily saturated which limits parallel heating by the shock potential and magnetic reconnection, and the slow growth of the electron/ion temperature ratio in our model would not necessarily continue. For $20<\ma<100$ our model predicts that the temperature ratio} is independent of the shock velocity. We do not reproduce in our simulations the the empirical relation $\te/\ti \propto \vsh^{-2}$ \citep{2008ApJ...689.1089V}, which for the standard jump conditions for $\ti$ is equivalent to stating that $\te$ is independent of $\vsh$ and hence electron heating not governed by the bulk-kinetic energy supply of inbound particles. 
\ab{Instead, we see that the electron temperature is roughly proportional to the ion upstream kinetic energy.} \mpo{Caution is advised when comparing the temperature ratio immediately downstream of the shock, that is measured in our simulations, with that found in observations which reflects the state of plasma a few months or years after passage through the shock.}

\ab{The Solar system also hosts high Mach number shocks. One of the} \pjm{best studied} \ab{examples is Saturn's bow shock whose properties have been measured in-situ by the Cassini spacecraft. The plasma beta upstream of Saturn's bow shock is about 0.1 \citep{2011SoPh..274..481J}, which is within the parameter range of our simulations. The postshock temperature ratio for the quasi-perpendicular Saturn's bow shock was published by \cite{2011JGRA..11610107M}.  Figure~\ref{total-heating-saturn} shows a comparison of the heating model with these data. For comparison, we consider cases when the shock} \mpo{has a high Mach number ($\ma > 20$) and is quasi-perpendicular ($\theta_{B_n} > 45^o$), which severely limits the number of data points. Cassini can only measure the electron temperature. The ion temperature is inferred from the shock speed that itself is an estimate requiring transformation from the spacecraft frame to the shock frame. \mpo{To indicate the level of correction that \cite{2011JGRA..11610107M} had to impose, we add to Fig.~\ref{total-heating-saturn} grey arrows that undo thoses changes and point to the original measurement.} Some data points are consistent within the errorbars provided by our model, but two} \pjm{values} \ab{are below the prediction. This may happen for few reasons. Our simulations consider only perpendicular shocks, but the \mpo{inferred shock obliquity for the shock crossings shown in Fig.~\ref{total-heating-saturn} is in the range $55^o < \theta_{B_n}<80^o$.} Second, in our simulations shocks propagate in homogeneous media, and we can not account for density or field} \mpo{fluctuations, both of which are present upstream of real shocks, that may impact on the temperature ratio}. \pjm{These assumptions therefore} \ab{lead to} \pjm{ additional uncertainties} \ab{in} \pjm{our} \ab{calculations of the temperature ratio. Nevertheless, we can state here that the heating model yields a \mpo{reasonably good} estimate of the temperature ratio at quasi-perpendicular high Mach number shocks.}

\acknowledgments

The authors thank Adam Masters and Ali Sulaiman for providing Saturn bow shock data. Great thanks to Aaron Tran, Lorenzo Sironi, Anatoly Spitkovsky and Vassilis Tsiolis for fruitful discussions. The work of J.N. has been supported by Narodowe Centrum Nauki through research project 2019/33/B/ST9/02569. This work was supported by JSPS-PAN Bilateral Joint Research Project Grant Number 180500000671. This research was supported by PLGrid Infrastructure. The numerical experiment was possible through a 10 Mcore-hour allocation on the 2.399 PFlop Prometheus system at ACC Cyfronet AGH. Part of the numerical work was conducted on resources provided by the North-German Supercomputing Alliance (HLRN) under projects bbp00014 and bbp00033.

\bibliographystyle{apj}
\bibliography{ref}

\end{document}